\documentclass[a4paper,11pt]{article}
\pdfoutput=1
\usepackage{jcappub}
\usepackage{mathtools}
\usepackage{ulem}
\usepackage{bm}
\usepackage{xcolor}

\renewcommand{\thesection}{\arabic{section}}

\def\laq{~\raise 0.4ex\hbox{$<$}\kern -0.8em\lower 0.62ex\hbox{$\sim$}~}
\def\gaq{~\raise 0.4ex\hbox{$>$}\kern -0.7em\lower 0.62ex\hbox{$\sim$}~}

\def\beq{\begin{equation}}
\def\eeq{\end{equation}}
\def\bea{\begin{eqnarray}}
\def\eea{\end{eqnarray}}

\def \pa {\partial}

\def \ti {\widetilde}

\def \ga {\gamma}

\def \Hcal {\mathcal{H}}
\def \U {\Upsilon}

\title{The Cosmological Perturbation Theory on the Geodesic Light-Cone background}

\author[a]{G. Fanizza,}
\author[b]{G. Marozzi,}
\author[b]{M. Medeiros,}
\author[b]{G. Schiaffino}

\affiliation[a]{Instituto de Astrofis\'ica e Ci\^encias do Espa\c{c}o, Faculdade de Ci\^encias da Universidade de Lisboa, Edificio C8, Campo Grande, P-1740-016, Lisbon, Portugal}

\affiliation[b]{Dipartimento di Fisica, Universit\`a di Pisa, Largo B. Pontecorvo 3, 56127 Pisa, 
Italy,\\
and Istituto Nazionale di Fisica Nucleare, Sezione di Pisa, Italy}

\emailAdd{gfanizza@fc.ul.pt}
\emailAdd{giovanni.marozzi@unipi.it}
\emailAdd{matheus.rodriguesmedeirossilva@phd.unipi.it}
\emailAdd{gloria.giorgia@hotmail.it}

\abstract{Inspired by the fully non-linear Geodesic Light-Cone (GLC) gauge, we consider its analogous set of coordinates which describes the unperturbed Universe.  Given this starting point, we then build a cosmological perturbation theory on top of it, study the gauge transformation properties related to this new set of perturbations and show the connection with standard cosmological perturbation theory. 
In particular, we obtain which gauge in standard perturbation theory corresponds to the GLC gauge, and put in evidence how this is a useful alternative to the standard Synchronous Gauge.
Moreover, we exploit several viable definitions for gauge invariant combinations. Among others, we build the gauge invariant variables such that their values equal the ones of linearized GLC gauge perturbations. This choice is motivated by two crucial properties of the GLC gauge: i) it admits simple expressions for light-like observables, e.g. redshift and angular distance, at fully non-linear level and ii) the GLC proper time coincides with the one of a 
free-falling observer. Thanks to the first property, exact expressions can then be easily expanded at linear order to obtain linear gauge invariant expression for the chosen observable. Moreover, the second feature naturally provides gauge invariant expressions for physical observables in terms of the time as measured by such free-falling observer. Finally, we explicitly show all these aspects for the case of the linearized angular distance-redshift relation.}

\keywords{
cosmological perturbation theory,
geodesic light-cone gauge
 
\vskip13pt plus8pt minus11pt

\noindent{\bfseries\large\sffamily{Preprints:}} 
}

\begin{document}

\maketitle

\section{Introduction}

One of the key aspects of observational cosmology is the fact that we observe the structures all across the Universe along our past-light cone, whereas we are so far able to describe the evolution of cosmological perturbations   on space-like hypersurfaces. The link between these two descriptions leads to our understanding of the observed Universe. However, connecting them is already a challenging task at linear level in perturbations theory. In this regard, for example, the linear angular distance-redshift relation has been widely studied in literature \cite{Sasaki:1987ad,Barausse:2005nf,Bonvin:2005ps,DiDio:2012bu,Biern:2016kys,Yoo:2016vne,Scaccabarozzi:2017ncm} and the agreement among the results has been achieved only after several discussions.

On top of that, the precision of forth-coming observations \cite{Abate:2012za,Amendola:2012ys,Aghamousa:2016zmz} requires to go beyond linear order to properly compare predictions with observations, such that non-linearities must be taken into account as well. In this view, also second order angular distance-redshift relation has been studied \cite{Gasperini:2011us,BenDayan:2012pp,BenDayan:2012ct,BenDayan:2012wi,Fanizza:2013doa,Marozzi:2014kua,Umeh:2014ana,Clarkson:2014pda}  and applied to investigate, for instance, the effects of the ensemble and direction averages on several distance-redshift relations \cite{BenDayan:2012pp,BenDayan:2012ct,BenDayan:2013gc,Bonvin:2015kea,Fleury:2016fda,Fanizza:2019pfp}. However,
 for what concerns non-linear angular distance-redshift relation a general agreement still needs to be achieved, especially on the role and presence of perturbations at the observer position.
 
Because of the above-mentioned reasons, having a theory for the evolution of cosmic structures directly given along our past light-cone we would significantly simplify their descriptions.
To this aim, we consider the so-called Geodesic Light-Cone (GLC) coordinates \cite{Gasperini:2011us} $x^\mu= (\tau, w, \ti \theta^a)$, $a=1,2$, where the most general cosmological metric can be parametrized in terms of the six arbitrary function $\U$, $U^a$, $\ga_{ab}= \ga_{ba}$, and the line element takes the form
\beq
ds^2_{GLC}=-2\U dwd\tau+\U^2dw^2+\ga_{ab}\left(d\ti\theta^a-U^a dw\right)\left(d\ti\theta^b-U^b dw\right).
\label{31}
\eeq
We recall that $w$ is a null coordinate, that photons travel along geodesics with constant $w$ and $\tilde\theta^a$, and that $\tau$ coincides with the time coordinate of the synchronous gauge \cite{BenDayan:2012pp}. The great advantage of the GLC line element in Eq. \eqref{31} consists of the fact that within this fully non-linear gauge choice, both time-like and light-like geodesic equations have exact solutions \cite{Gasperini:2011us,Fleury:2016htl}. In particular, the quadri-momentum of a light-like signal solves its geodesic equation as $k_\mu =\pa_\mu w$ whereas the quadri-velocity of a time-like observer chosen to correspond to the static free-falling observer in the Synchronous Gauge (SG) is given by $u_\mu =\pa_\mu \tau$.

Since the solution of light-like geodesics are available, this gauge shows extreme advantages in finding the fully non-linear expression of light-like observables such as redshift and angular (or luminosity) distance \cite{Gasperini:2011us,BenDayan:2012pp,BenDayan:2012ct,BenDayan:2012wi,Marozzi:2014kua,Fanizza:2013doa}, and galaxy number count \cite{DiDio:2014lka,DiDio:2015bua}. In particular, it turns out that the above-mentioned observables can be easily written as factorization\footnote{GLC gauge has shown also great advantages in the study of observables involving time-like observables in the ultra-relativistic regime. In this case, specific observables can be explicitly written fully non-linearly \cite{Fanizza:2015gdn} but not in terms of the factorized metric component.}  of the metric component evaluated at the observer and source position. So far, one problem of this gauge choice was that dynamics for a given model is hard to be solved, already at a linear level. In this regard, \cite{Mitsou:2020czr} provides a first attempt to find solution at such linear level by looking at the metric in Eq. \eqref{31} as a particular case of an ADM-like foliation, applied twice on a fully non-linear metric, where light-like geodesic are exactly solved but the observer motion is allowed to be non-geodesic. However, a lot of questions remains to be answered already in a linearized theory in such a context. In particular, as we will see below, one of the most important result in this work consists of showing how to formulate a gauge invariant perturbation theory which preserves the advantages of the non-linear GLC gauge and allows to be independent on the chosen gauge at the same time.

We will show that such a construction is possible and, for the chosen example of the angular distance-redshift relation, we will provide a general gauge invariant expression where the time at the observer is chosen to coincide with the time of a free-falling observer which is static in the SG. The expression for the angular distance that we are about to obtain corresponds to the solution one would get by solving the angular distance expression through the Jacobi map approach. We will then show that the linearized GLC gauge is the only one among the others, where observables can be written as a sum of perturbations at the observer and at the source position.

We underline that the perturbative theory discussed in this paper is different from what has been done in \cite{Mitsou:2020czr} for the following two aspects. First of all, in \cite{Mitsou:2020czr} perturbations are gauge fixed to describe the past light-cone of the observer, differently from what we will show here. Secondly, along this work we fix the background to the coordinates described by an observer comoving with a geodesic time-like flow. In \cite{Mitsou:2020czr} this assumption is not required neither in the exact equations nor in the perturbative ones. This means that the dynamical equations studied in \cite{Mitsou:2020czr} also at the background level admit observers to be not free falling. This generalization might be well-suited for the description of the cosmological eras where the pressure is not constant, e.g. during the radiation domination eras \cite{Kodama:1985bj}. However, this aspect goes beyond the purposes of this paper since, as a general framework, we are interested in the study of Large Scale Structure, hence well after the decoupling time. We leave the detailed discussion of this generalization to future studies devoted to that. We remark that for what concerns the other degrees of freedom, \cite{Mitsou:2020czr} strictly works within the Light-Cone fixing of the GLC gauge whereas in this work we start from a general set of perturbations in order to find a well-suited set of general gauge invariant variables.

The paper is structured as follows. In Sect. \ref{sec:2}, we firstly recall the standard linear perturbation theory and then show how to build linear perturbation theory on the background geodesic light-cone coordinates. Then we derive the gauge transformation properties within the new light-cone framework and discuss the analogies with the standard theory. Moreover, we provide a classification among all the viable choices for the identification of gauge invariant variables and discuss few cases of physical interest.
In Sect. \ref{sec:3}, we start from the procedure outlined in Sect. \ref{sec:2} and derive the gauge invariant variables in terms of the light-cone perturbations such that their values equal the perturbations of the linearized GLC gauge. After that, we also discuss the link between the possible gauge choices in standard perturbation theory and in the new proposed one.
Hence, in Sect. \ref{sec:4} we firstly show that the GLC gauge is the one well-suited to evaluate gauge invariant physical observables in terms of the proper time as measured by a free-falling observer which is static in the SG. This result agrees with, and generalizes to any light-like observable, what has been shown in literature for the angular distance-redshift relation (see \cite{Biern:2016kys}). We then identify a general procedure to derive gauge invariant expression for the light-like observable in terms of the light-cone perturbations which preserves the advantages of the GLC gauge in the technical evaluations.
In Sect. \ref{sec:5}, we then apply the above-mentioned prescription to the case of the angular distance-redshift relation and find its expression in terms of all the light-like perturbations, regardless of the chosen gauge adopted. The expression that we find turns out to be gauge invariant as expected both for the observer and source perturbations.
In Sect. \ref{sec:6}, we explicitly check that our linear gauge invariant result for the angular distance-redshift relation remains indeed gauge invariant even when written in terms of the standard perturbation theory. Moreover, we compare our results with the one already presented in literature.
Finally, in Sect. \ref{sec:7} we discuss and summarize our main results and argue about future applications.
To conclude, in Appendix \ref{app:w_o}, we discuss the residual gauge fixing of the GLC gauge invariant variables and its meaning in terms of standard perturbations.

\section{Cosmological Perturbation Theory on the background Geodesic Light-Cone coordinates}
\label{sec:2}

Let us start from the perturbed FLRW metric, expressed as function of the conformal time $\eta$ and spherical coordinates, $y^\mu=\left(\eta,r,\theta^a \right)$,
\begin{align}
ds^2=&g_{\mu\nu}dy^\mu dy^\nu
=\left(\bar{g}_{\mu\nu}+\delta g_{\mu\nu}\right)dy^\mu dy^\nu
\nonumber\\
=&\,a^2(\eta)\left[ -\left(1+2\phi\right)d\eta^2
-2\,\mathcal{B}_rdrd\eta
-2\,\mathcal{B}_ad\theta^ad\eta
+\left(1+\mathcal{C}_{rr}\right)dr^2\right.
\nonumber\\
&\left.+\left(\bar{\gamma}^{FRW}_{ab}+\mathcal{C}_{ab}\right)d\theta^ad\theta^b
+2\,\mathcal{C}_{ra}drd\theta^a \right]\,.
\label{eq:FRWpert}
\end{align}
Here $\bar{g}_{\mu\nu}$ represents the background metric, $\delta g_{\mu\nu}$ are the linear perturbations and $\bar{\gamma}^{FRW}_{ab}=\text{diag}\left(r^{2}, r^{2}\sin^{2}\theta\right)$, where $a=1,2$ refers to the angular coordinates $\left( \theta,\phi \right)$. In Eqs. \eqref{eq:FRWpert} the perturbations can be decomposed accordingly to their transformation properties with respect to the background symmetry $SO(3)$ as follow
\begin{align}
\mathcal{B}_{r}=&\,\partial_{r}B+B_{r}\qquad,\qquad\mathcal{B}_{a}=\,\partial_{a}B+B_{a}\,,\nonumber\\
\mathcal{C}_{rr}=&-2\,\psi+2\,\bar{D}_{rr}E+2\nabla_{r}F_{r}+2h_{rr}\,,\nonumber\\
\mathcal{C}_{ab}=&-2\,\psi\,\bar{\gamma}^{FRW}_{ab}+2\,\bar{D}_{ab}E+2\nabla_{(a}F_{b)}+2h_{ab}\,,\nonumber\\
\mathcal{C}_{ra}=&2\,\bar{D}_{ra}E+2\nabla_{(r}F_{a)}+2h_{ra}\,,
\label{eq:FLRW_dec}
\end{align}
where $\phi,\,\psi,\,E$ and $B$ are scalars. Furthermore, considering an index $i=(r, \theta, \phi)$ which to spatial polar coordinates, in the above equations we have that $B_i$ and $F_i$ are divergenceless vectors and $h_{ij}$ is a traceless and divergenceless tensor. According to that $\nabla_i$ are the covariant derivatives and $\bar{D}_{ij}\equiv\nabla_{(i}\nabla_{j)}-\frac{1}{3}a^{-2}\bar{g}_{ij}\Delta_3$ is a traceless operator such that the entire trace of the spatial metric is addressed to $\psi$ and $\Delta_3$ is the three-dimensional Laplacian.

The background metric in Eq. \eqref{eq:FRWpert} can be equivalently expressed in terms of the background geodesic light-cone coordinates $x^\mu=\left( \tau,w,\tilde\theta^a \right)$ where, as mentioned above, $\tau$ is the proper time of a free-falling time-like particle, $w$ describes the background light-cone and $\tilde{\theta}^a$ stands for the angular coordinates and coincides with $\theta^a$ on the background. Thanks to a finite coordinate transformation, $x^\mu$ and $y^\mu$ are trivially related as
\begin{equation}
\eta(\tau)=\int_{\tau_{in}}^\tau\frac{d\tau'}{a(\tau')}\qquad,\qquad
r=w-\eta(\tau)\qquad,\qquad
\theta^a=\tilde\theta^a\,.
\label{eq:finiteCT}
\end{equation}
According to these relations, since the line element is invariant under coordinate transformation, the metric in Eq. \eqref{eq:FRWpert} transforms as
\begin{equation}
f_{\mu\nu}=\frac{\partial y^{\alpha}}{\partial x^{\mu}}\frac{\partial y^{\beta}}{\partial x^{\nu}}g_{\alpha\beta},
\label{eq:24}
\end{equation}
leading to
\begin{align}
ds^2=&f_{\mu\nu}dx^\mu dx^\nu
=\left(\bar{f}_{\mu\nu}+\delta f_{\mu\nu}\right)dx^\mu dx^\nu
\nonumber\\
=&a^2(\tau)\left[ L\,d\tau^2
-\frac{2}{a}\left(1-a\,M \right)d\tau dw
+2\,V_bd\tilde\theta^bd\tau 
+\left( 1+N \right)dw^2\right.
\nonumber\\
&\left. +2\,U_ad\tilde\theta^a dw
+\left(\bar{\gamma}_{ab}+\delta\gamma_{ab}\right)d\tilde\theta^ad\tilde\theta^b\right]\,,
\label{eq:GLCpert}
\end{align}
where
\begin{equation}
\bar{f}_{\mu\nu}=\frac{\partial y^{\alpha}}{\partial x^{\mu}}\frac{\partial y^{\beta}}{\partial x^{\nu}}\bar{g}_{\alpha\beta}
=a^2(\tau)\left[ 
-\frac{2}{a}d\tau dw
+dw^2
+\bar{\gamma}_{ab}d\tilde\theta^ad\tilde\theta^b\right]
\end{equation}
is the background metric in terms of the geodesic light-cone coordinates, $\delta f_{\mu\nu}\equiv f_{\mu\nu}-\bar{f}_{\mu\nu}$ are the linear perturbations on top of $\bar{f}_{\mu\nu}$, linked to $\delta g_{\mu\nu}$ through Eqs. \eqref{eq:finiteCT} and \eqref{eq:24} as
\begin{equation}
\delta f_{\mu\nu}=\frac{\partial y^{\alpha}}{\partial x^{\mu}}\frac{\partial y^{\beta}}{\partial x^{\nu}}\delta g_{\alpha\beta}
\end{equation}
and $\bar{\gamma}_{ab}=\left[ w-\eta(\tau) \right]^2\text{diag}\left( 1,\sin^2\tilde\theta^1 \right)\equiv \left[ w-\eta(\tau) \right]^2 q_{ab}$, with $a,b=1,2$.

The background in Eq. \eqref{eq:GLCpert} shares the same symmetries of $\bar{g}_{\mu\nu}$. This means that perturbations in $\delta f_{\mu\nu}$ can be classified in terms of the same group of symmetries as done for $\delta g_{\mu\nu}$. In particular, $V_a,U_a$ and $\gamma_{ab}$ can be decomposed according to their transformations laws with respect to rotations on the unitary sphere $q_{ab}$. In this way (see also \cite{Schiaffino2019,Mitsou:2020czr}) we consider the covariant derivative $D_a$ on the unitary sphere and its dual $\widetilde{D}_{a}=\epsilon_{a}^{b}D_{b}$, with $\epsilon_{a}^{b}$ anti-symmetric tensor, and then decompose our perturbations as
\begin{align}
V_{a}=&r^{2}\left(D_{a}v+\widetilde{D}_{a}\hat{v}\right),\nonumber\\
U_{a}=&r^{2}\left(D_{a}u+\widetilde{D}_{a}\hat{u}\right),\nonumber\\
\delta\gamma_{ab}=&2\,r^{2}\left[q_{ab}\nu+D_{ab}\mu+\widetilde{D}_{ab}\hat{\mu}\right]\,,
\label{eq:GLC_dec}
\end{align}
where $D_{ab}=D_{(a}D_{b)}-\frac{1}{2}q_{ab}D^{2}$, $D^{2}$ is the angular laplacian and $\widetilde{D}_{ab}=D_{(a}\widetilde{D}_{b)}$. Since $\delta\gamma_{ab}$ is a symmetric $2\times 2$ matrix, we have decoupled its 3 degrees of freedom as its trace $\nu$ and its traceless counterparts $\mu$ and $\hat{\mu}$. Analogously, since $U_a$ and $V_a$ are 2 dimensional vectors, they can always be decomposed as their divergence $v$ and $u$ and their divergence-less degrees of freedom $\hat{v}$ and $\hat{u}$ (see \cite{Mitsou:2019nhj} for a general decomposition of perturbations on the 2-sphere). Thanks to this classification, we end up with 10 degrees of freedom as expected: 7 of them, i.e. $L,\,M,\,N,\,u,\,v,\,\mu$ and $\nu$ transform as scalar quantity under rotations on the unitary sphere, whereas the remaining 3, namely $\hat{u},\,\hat{v}$ and $\hat{\mu}$, transforms as pseudo-scalar under the same symmetry group.

Clearly, not all these degrees of freedom are physical. Indeed we can eliminate 4 of them thanks to a gauge choice. In the following, we will show that these 6 physical degrees of freedom consist of 4 scalars and 2 pseudo-scalars.

\subsection{Gauge invariant variables: standard and light-cone perturbations}
Before discussing the gauge choice for the perturbations in $\delta f_{\mu\nu}$, let us review the gauge fixing in standard cosmological perturbations theory. Under a linear shift of the coordinates $y^\mu \rightarrow \tilde y^\mu=y^\mu+\epsilon^\mu$, $\delta g_{\mu\nu}$ transforms as
\begin{equation}
\delta \widetilde{g_{\mu\nu}}(y^\mu)= \delta g_{\mu\nu}(y^\mu)-\nabla_{\mu}\epsilon_{\nu}-\nabla_{\nu}\epsilon_{\mu}\,,
\label{eq:GT}
\end{equation}
where $\nabla_\mu$ is the covariant derivative and we stress that both l.h.s. and r.h.s. of \eqref{eq:GT} are evaluated at the same background coordinates $y^\mu$.
In this way, the metric perturbations transform as
\begin{align}
\widetilde{\phi}=&\phi-\Hcal \epsilon^\eta-\partial_\eta\epsilon^\eta,\nonumber\\
\widetilde{\mathcal{B}_{r}}=&\mathcal{B}_{r}-\partial_{r}\epsilon^\eta+\partial_\eta\epsilon^r,\nonumber\\
\widetilde{\mathcal{B}_{a}}=&\mathcal{B}_{a}-\partial_{a}\epsilon^\eta+\bar{\gamma}^{FRW}_{ab}\partial_\eta\epsilon^{b},\nonumber\\\widetilde{\mathcal{C}_{rr}}=&\mathcal{C}_{rr}-2\Hcal \epsilon^\eta-2\partial_{r}\epsilon^r,\nonumber\\
\widetilde{\mathcal{C}_{ra}}=&\mathcal{C}_{ra}-\partial_{a}\epsilon^r-\bar{\gamma}^{FRW}_{ab}\partial_{r}\epsilon^{b},\nonumber\\
\widetilde{\mathcal{C}_{ab}}=&\mathcal{C}_{ab}-\frac{1}{a^{2}}\epsilon^{\rho}\partial_{\rho}\left(a^{2}\bar{\gamma}^{FRW}_{ab}\right)-(\bar{\gamma}^{FRW}_{ac}\partial_{b}+\bar{\gamma}^{FRW}_{bc}\partial_{a})\epsilon^{c},
\label{eq:FLRW_gauge}
\end{align}
where $\Hcal=\partial_\eta a/a$. Similarly to what done in Eq. \eqref{eq:FLRW_dec}, the gauge field too can be decomposed as 
$\epsilon^i=\partial^i\epsilon+e^i$, where $\nabla_i e^i=0$. This clearly separates the spatial gauge modes as a scalar one $\epsilon$ and a divergenceless one $e^i$. Since no-tensor gauge degree of freedom are present, it follows that $h_{ij}$ is gauge invariant, as well-known. For illustrative purposes, we also recall a possible way to build gauge invariant variables for scalar and vector perturbations in standard perturbations theory. Combining Eqs. \eqref{eq:FLRW_gauge} and \eqref{eq:FLRW_dec}, we get the well-known transformation's rules for the scalars (see, for example, \cite{Mukhanov:1990me})
\begin{equation}
\widetilde{\psi}=\psi+\Hcal\,\epsilon^\eta+\frac{1}{3}\Delta_3\epsilon\quad,\quad
\widetilde{\phi}=\phi-\partial_\eta\epsilon^\eta-\Hcal\,\epsilon^\eta\quad,\quad
\widetilde{E}=E-\epsilon\quad,\quad
\widetilde{B}=B-\epsilon^\eta+\partial_\eta\epsilon\,.
\label{eq:FLRW_dec_GT}
\end{equation}
It is helpful for our purposes to underline that we can define gauge invariant variables to be equal to the value taken by our perturbations in a specific gauge. For example, if we work within the scalar sector of linear perturbations and refer to the Longitudinal Gauge (LG), where $\widetilde E=\widetilde B=0$, the infinitesimal coordinates transformation generator $\epsilon^\mu$ takes the value    
\begin{equation}
\epsilon^\eta=B+\partial_\eta E\qquad\text{and}\qquad
\epsilon=E\,
\label{eq:210}
\end{equation}
and the corresponding gauge invariant variables are
\begin{equation}
\Psi_{LG}=\psi+\mathcal{H}\left( B+\partial_\eta E \right)+\frac{1}{3}\Delta_3 E\qquad,\qquad
\Phi_{LG}=\phi-\partial_\eta\left( B+\partial_\eta E \right)-\mathcal{H}\left( B+\partial_\eta E \right)\,,
\label{eq:Bardeen}
\end{equation}
which are nothing but the so-called Bardeen potentials. On the other hand, also different choices are allowed. As an instance, we can identify the value of the gauge invariant variables as given by the value of linear perturbations in the Uniform Curvature Gauge (UCG), where 
$\widetilde\psi=\widetilde E=0$. From Eqs. \eqref{eq:FLRW_dec_GT}, this leads to
\begin{equation}
\epsilon^\eta=-\frac{\psi}{\mathcal{H}}-\frac{\Delta_3 E}{3\Hcal}\qquad\text{and}\qquad
\epsilon=E
\end{equation}
and
\begin{equation}
\Phi_{UCG}=\phi+\partial_\eta\left({\frac{\psi}{\mathcal{H}}+\frac{\Delta_3 E}{3\Hcal}}\right)+\psi+\frac{\Delta_3 E}{3\Hcal}\qquad,\qquad
\mathcal{B}_{UCG}=B+\frac{\psi}{\mathcal{H}}+\frac{\Delta_3 E}{3\Hcal}+\partial_\eta E\,.
\label{eq:GI_UCG}
\end{equation}
The choices made for the variables \eqref{eq:Bardeen} and \eqref{eq:GI_UCG} completely fix the potentials and no residual gauge freedom is present. This is due to the fact that for such choices the conditions imposed on the gauge fields are imposed directly on the fields themselves. Other choices, as for example identifying the gauge invariant variables with the perturbations within the SG, are viable. However, in this latter case only the time derivatives of the gauge fields can be fixed. We can immediately realize that by imposing $\widetilde \phi=\widetilde B=0$ in Eqs. \eqref{eq:FLRW_dec_GT}. This leads to
\begin{equation}
\partial_\eta\left(a\epsilon^\eta\right)=a\phi\qquad\text{and}\qquad
\partial_\eta\epsilon=B+\epsilon^\eta.
\label{eq:214}
\end{equation}
This means that the obtained gauge invariant combinations contains a symmetry under a time-independent shift of $\epsilon$.

The same procedure can be applied to vector perturbations. They transform as
\beq
\widetilde{B_i}=B_i+\pa_\eta\left( \frac{e_i}{a^2} \right)\qquad\text{and}\qquad
\widetilde{F_i}=F_i-\frac{e_i}{a^2}\,.
\eeq
We can then require that the vector gauge invariant variable $\Psi_i$ takes the value of the vector perturbation $B_i$ in the gauge where $\widetilde{F_i}=0$. In this case we have $e_i=a^2F_i$ and obtain
\beq
\Psi_i=B_i+\pa_\eta F_i\,.
\eeq
For the vector perturbations let us note that this choice is the unique that does not involve derivatives of $e^i$, so that with this choice $\Psi_i$ is completely fixed.

The same procedure can be followed for the perturbations $\delta f_{\mu\nu}$. Indeed, even in this case we have that under a linear shift $x^\mu\rightarrow \tilde{x}^\mu=x^\mu+\xi^\mu$, the metric perturbation $\delta f_{\mu}$ transforms as
\begin{equation}
\delta \widetilde{f_{\mu\nu}}(x^\mu)= \delta f_{\mu\nu}(x^\mu)-\nabla_{\mu}\xi_{\nu}-\nabla_{\nu}\xi_{\mu}\,,
\label{eq:f_GT}
\end{equation}
where now $\xi^\mu=\left( \xi^0,\xi^w,\hat{\xi}^a \right)$. This leads to the following gauge transformation for the metric perturbations in Eq. \eqref{eq:GLCpert}
\begin{align}
\widetilde{L}=&L+\frac{2}{a}\,\dot{\xi}^w,\nonumber\\
\widetilde{M}=&M+\frac{1}{a}H\xi^0-\dot{\xi}^w+\frac{1}{a}\left({\xi^w}'+\dot{\xi}^0\right),\nonumber\\
\widetilde{N}=&N-2H\xi^0+\frac{2}{a}{\xi^0}'-2{\xi^w}',\nonumber\\
\widetilde{V_{a}}=&V_{a}+\frac{1}{a}\partial_{a}\xi^w-\bar{\gamma}_{ab}\dot{\hat{\xi}}^{b},\nonumber\\
\widetilde{U_{a}}=&U_{a}+\frac{1}{a}\partial_{a}\xi^0-\partial_{a}\xi^w-\bar{\gamma}_{ab}\,\hat{\xi}^b\,',\nonumber\\
\widetilde{\delta\gamma_{ab}}
=&\delta\gamma_{ab}
-\frac{1}{a^{2}}\xi^0\left(a^{2}\bar\gamma_{ab}\right)\dot{}
-\xi^w\left(\bar\gamma_{ab}\right)'
-\left(\bar\gamma_{ac}D_{b}+\bar\gamma_{bc}D_{a}\right)\hat{\xi}^{c},
\label{eq:GLC_gauge}
\end{align}
where $\dot{ }\equiv\partial_\tau$, $'\equiv\partial_w$ and $H=\dot{a}/a$. The infinitesimal coordinate transformation generators $\hat{\xi}^a$ in Eqs. \eqref{eq:GLC_gauge} can be again decomposed according to its transformation properties for any rotation on the unitary sphere\footnote{This is the analogous of standard cosmological perturbation theory, when linear diffeomorphism is decomposed in the same way as vector perturbations.}. Therefore, we can write
\begin{equation}
\hat{\xi}^a=q^{ab}\left(D_b\,\chi+\widetilde{D}_b\,\hat{\chi}\right)\,,
\label{eq:chi_dec}
\end{equation}
where $\chi$ is a scalar gauge degree of freedom, as well as $\xi^0$ and $\xi^w$, and $\hat{\chi}$ is the only pseudo-scalar degree of freedom. Thanks to this further classification, the transformation laws for the 10 degrees of freedom in $\delta f_{\mu\nu}$ become
\begin{align}
\widetilde{L}=&L+\frac{2}{a}\dot{\xi^w},\nonumber\\
\widetilde{M}=&M+\frac{1}{a}H\xi^0-\dot{\xi^w}+\frac{1}{a}\left({\xi^w}'+\dot{\xi^0}\right),\nonumber\\
\widetilde{N}=&N-2H\xi^0+\frac{2}{a}{\xi^0}'-2{\xi^w}',\nonumber\\
\widetilde{\nu}=&\nu-\frac{1}{2}D^{2}\chi-\xi^0 \left( H-\frac{1}{ar} \right)-\frac{\xi^w}{r},\nonumber\\\widetilde{\mu}=&\mu-\chi\qquad,\qquad
\widetilde{\hat{\mu}}=\hat{\mu}-\hat{\chi}\nonumber\\
\widetilde{v}=&v +\frac{1}{ar^{2}}\xi^w-\dot{\chi}\qquad,\qquad
\widetilde{u}=u +\frac{1}{ar^{2}}\xi^0-\frac{\xi^w}{r^{2}}-\chi',\nonumber\\
\widetilde{\hat{v}}=&\hat{v}-\dot{\hat{\chi}}\qquad,\qquad
\widetilde{\hat{u}}=\hat{u}-\hat{\chi}'.
\label{eq:dec_GLC_GT}
\end{align}
At this point, we define \textit{algebraic fixing} the choice for the value of the gauge invariant variables to coincide with the perturbations in a gauge with no left residual gauge freedom. On the other hand, we call \textit{analytic fixing} the opposite case where a residual gauge freedom is still present in the choice of the gauge invariant variable. This nomenclature reflects the fact that in the former case, one entirely fixes the gauge modes whereas in the latter case one imposes condition on the derivatives of the linear gauge modes. This classification can be better understood by referring to the standard cosmological perturbation theory discussed at the beginning of this subsection. Indeed the Bardeen variables in Eqs. \eqref{eq:Bardeen} regard an algebraic fixing, since they correspond to the perturbations in the LG, where no residual gauge freedom is left, as explicitly shown in Eqs. \eqref{eq:210}. On the contrary, the gauge invariant variables chosen to coincide with the SG perturbations corresponds to an analytic fixing, since in that case one only imposes a condition on the time derivative of the spatial gauge mode, as shown in Eq. \eqref{eq:214}.

Applying this classification to the cosmological perturbation theory on the geodesic light-cone background defined above, we note that several algebraic fixing are possible. However, from Eqs. \eqref{eq:dec_GLC_GT} any algebraic fixing must require $\mu=\hat\mu=0$ to completely fix $\chi$ and $\hat\chi$, since these gauge transformations are the only ones which involve $\chi$ and $\hat\chi$ without their derivatives. Let us also note that the GLC gauge fixing does not belong to this latter case. First of all because GLC gauge is an analytic fixing, since it admits residual gauge freedom, and secondly because this residual gauge freedom also involves $\chi$ and $\hat\chi$. We will show in the following sections how to require the analytic fixing corresponding to the linearized GLC gauge and obtain the correspondent gauge invariant variables.

In the last part of this section, we just list and discuss few interesting choices for the algebraic fixing.

\subsubsection{Magnetic Gauge}
We can identify a set of gauge invariant variables such that they correspond to the perturbations in the gauge where $\widetilde\mu=\widetilde{\hat\mu}=\widetilde u=\widetilde v=0$. In this case, $U_a$ and $V_a$ are entirely sourced by pseudo-scalar variables, so electric and magnetic part of the variables are completely separated in the metric tensor. We refer to this choice as the \textit{magnetic gauge}. The reason for this name stands in the fact that this gauge choice completely decouples in the metric components the scalar degrees of freedom (or \textit{electric} ones) from the pseudo-scalar ones (or \textit{magnetic} ones). Indeed, in general such a feature is not true, since $U_a$, $V_a$ and $\delta\gamma_{ab}$ is general can be sourced by both scalar and pseudoscalar degrees of freedom. Among all the algebraic fixings, this gauge is the unique ones providing such a decoupling. In this case, the gauge invariant variables can be obtained from Eqs. \eqref{eq:dec_GLC_GT} to be
\begin{align}
\mathbb{L}_m=&L+\frac{2}{a}\left[ar^2\left(\dot{\mu}-v\right)\right]\dot{}\,\,,\nonumber\\
\mathbb{M}_m=&M+\frac{1}{a}H\,\mathbb{T}_m-\left[ar^2\left(\dot{\mu}-v\right)\right]\dot{}+\frac{1}{a}\left\{\left[ar^2\left(\dot{\mu}-v\right)\right]'+\dot{\mathbb{T}}_m\right\},\nonumber\\
\mathbb{N}_m=&N-2H\,\mathbb{T}_m+\frac{2}{a}\mathbb{T}_m'-2\left[ar^2\left(\dot{\mu}-v\right)\right]',\nonumber\\
\mathcal{V}_m=&\nu-\frac{1}{2}D^2\mu-\mathbb{T}_m \left( H-\frac{1}{ar} \right)
-ar\left(\dot{\mu}-v\right),\nonumber\\
\hat{\mathbb{V}}_m=&\hat{v}-\dot{\hat{\mu}},\nonumber\\
\hat{\mathbb{U}}_m=&\hat{u}-\hat{\mu}'.
\end{align}
where $\mathbb{T}_m=-ar^2\left(u -\mu'+av-a\dot{\mu}\right)$ and the subscript $m$ stands for magnetic gauge.

\subsubsection{Spherical V Gauge}

Another interesting choice consists of identifying the gauge invariant variables as the perturbations in the gauge with $\widetilde{\delta\gamma_{ab}}=0$ and $\widetilde v=0$. In this case there are no perturbations on the induces metric on the 2-sphere, so the angular metric is trivially $a^2r^2\,\text{diag}\left( 1,\sin^2\tilde\theta^1 \right)$ everywhere. Moreover, we have that only $V_a$ is entirely sourced by a pseudo-scalar field. This choice is interesting since leaves the metric on the 2-sphere completely unperturbed. For such a reason, we call it the \textit{spherical V gauge}. In this case, again from 
Eqs. \eqref{eq:dec_GLC_GT}, the gauge invariant variables are
\begin{align}
\mathbb{L}_V=&L+\frac{2}{a}\dot{\mathbb{W}}_V\,\,,\nonumber\\
\mathbb{M}_V=&M+\frac{1}{a}H\,\mathbb{T}_V-\dot{\mathbb{W}}_V+\frac{1}{a}\left\{\mathbb{W}_V'+\dot{\mathbb{T}}_V\right\},\nonumber\\
\mathbb{N}_V=&N-2H\,\mathbb{T}_V+\frac{2}{a}\mathbb{T}_V'-2\mathbb{W}_V',\nonumber\\
\mathbb{U}_V=&u +\frac{1}{ar^{2}}\mathbb{T}_V-\frac{1}{r^2}\mathbb{W}_V-\mu',\nonumber\\
\hat{\mathbb{V}}_V=&\hat{v}-\dot{\hat{\mu}},\nonumber\\
\hat{\mathbb{U}}_V=&\hat{u}-\hat{\mu}',
\label{eq:GLC_GI}
\end{align}
where we have defined $\mathbb{T}_V =\frac{ar}{1-aHr}\left[ar\left(\dot{\mu}-v\right)-\nu+\frac{1}{2}D^2\mu\right]$, $\mathbb{W}_V=ar^2\left( \dot{\mu}-v \right)$ and the subscript $V$ reminds that we are working in the spherical V gauge.

\subsubsection{Spherical U Gauge}

Just as the previous case, we can also define the \textit{spherical U gauge} where $\widetilde{\delta\gamma_{ab}}=0$ and $\widetilde u=0$ and then identify the gauge invariant variable as the perturbations in this gauge. This case is specular to the spherical V gauge since it addresses only pseudo-scalar field to $U_a$ rather than $V_a$. With this choice, we obtain
\begin{align}
\mathbb{L}_U=&L+\frac{2}{a}\dot{\mathbb{W}}_U,\nonumber\\
\mathbb{M}_U=&M+\frac{1}{a}H\mathbb{T}_U-\dot{\mathbb{W}}_U+\frac{1}{a}\left(\mathbb{W}_U'+\dot{\mathbb{T}}_U\right),\nonumber\\
\mathbb{N}_U=&N-2H\mathbb{T}_U+\frac{2}{a}\mathbb{T}_U'-2\mathbb{W}_U',\nonumber\\
\mathbb{V}_U=&v +\frac{1}{ar^{2}}\mathbb{W}_U-\dot{\mu},\nonumber\\
\hat{\mathbb{V}}_U=&\hat{v}-\dot{\hat{\mu}},\nonumber\\
\hat{\mathbb{U}}_U=&\hat{u}-\hat{\mu}'.
\end{align}
where $\mathbb{T}_U=-\frac{1}{H}\left(r u-r\mu'-\nu+\frac{1}{2}D^2\mu\right)$, $\mathbb{W}_U=r^2 u +\frac{1}{a}\mathbb{T}_U-r^2\mu'$ and the subscript recalls that we refer to the spherical U gauge.

Despite of the fact that all the algebraic fixing univocally select $\chi$ and $\hat\chi$, the fixing of $\xi^0$ and $\xi^w$ can be required algebraically in several way. The three above-mentioned fixing are the simplest ones, but we might also choose combinations of perturbations to be zero in the scalar sector.\footnote{For instance, from Eqs. \eqref{eq:dec_GLC_GT} we could impose $u+av=0$ to fix $\xi^0$ and then fix $\nu=0$ for $\xi^w$.} However, regardless of the choices for $\xi^0$ and $\xi^w$, we stress again that the two magnetic gauge invariant variables $\hat{\mathbb{U}}$ and $\hat{\mathbb{V}}$ always involve the same combination of perturbations. About the linearized GLC gauge, where (as shown below in Eq. \eqref{eq:GLC_linear_cond}) $V_a=0$, we then conclude that the combinations $-\dot{\hat{\mu}}$ and $\hat{u}-\hat{\mu}'$ corresponds to the gauge invariant variables for the magnetic part of the vector and tensor perturbations. Hence, their study is crucial in case one is interested in studying the magnetic part of linear gravitational waves.

As anticipated in the previous section, we note that only one pseudo-scalar degree of freedom can be eliminated. This fact can be understood by looking at the analogous in the standard perturbation theory. Indeed, also in that case there are only two pseudo-scalar degrees of freedom, i.e. the magnetic part of the vector and tensor gauge invariant variable. Since the physics is independent on the coordinate choice, we infer that it must be possible to define two gauge invariant pseudo-scalar variables such that one of those corresponds to the magnetic part of the gravitational waves.

The same interpretation is viable for the scalar degrees of freedom. In this case, we are left with 4 of them. Whereas two of them may be understood as analogous of the Bardeen variables, the remaining two must correspond to the electric part of the gauge invariant vector and tensor variables.

\section{Geodesic Light-Cone gauge invariant variables}
\label{sec:3}
So far, we have discussed the general treatment of the perturbations on a background described by geodesic light-cone coordinates. In this regard, we have shown how to obtain gauge invariant variables thanks to what we called the algebraic fixing. In the following, we want instead get a set of gauge invariant variables which coincides with the value that perturbations take in the linearized version of the GLC gauge of Eq. \eqref{31}. Such linearized version corresponds to the conditions
\begin{equation}
L=0\qquad,\qquad N+2aM=0\qquad\text{and}\qquad V_a=0\,.
\label{eq:GLC_linear_cond}
\end{equation}
We can then impose the conditions $\widetilde{L}=\widetilde{N}+2a\widetilde{M}=\widetilde{v}=\widetilde{\hat{v}}=0$  on the Eqs. \eqref{eq:dec_GLC_GT} and obtain
\begin{align}
\dot{\xi^0}+\frac{{\xi^0}'}{a}=-\frac{1}{2}\left(N+2aM+a^2 L\right)\qquad,&\qquad
\dot{\xi^w}=-\frac{a}{2}L,\nonumber\\
\dot{\chi}=v +\frac{1}{ar^{2}}\xi^w\qquad,&\qquad
\dot{\hat{\chi}}=\hat{v}\,.
\label{eq:GLC_fixing}
\end{align}
Let us underline that the choice of this gauge requires conditions only on the derivatives of the gauge modes. According to our classification then this corresponds to an analytic fixing. The solutions of Eqs. \eqref{eq:GLC_fixing} are then
\begin{align}
\xi^0=&-\frac{1}{2}\int^\tau_{\tau_{in}}d\tau'\left(N+2aM+a^2\,L\right)(\tau',w-\eta(\tau)+\eta(\tau')),\nonumber\\
\xi^w=&\frac{1}{2}\int^{\tau_o}_\tau d\tau'\,a L+w_o,\nonumber\\
\chi=&-\int^{\tau_o}_\tau d\tau' \left(v +\frac{1}{2\,ar^{2}}\int^{\tau_o}_{\tau'} d\tau''\,a L+\frac{w_o}{ar^2}\right)+\chi_o,\nonumber\\
\hat{\chi}=&-\int^{\tau_o}_\tau d\tau'\,\hat{v}+\hat\chi_o\,,
\label{eq:33}
\end{align}
where $\tau_{in}$ is an early enough time when perturbations (or better the integrand) were negligible, $\tau_o$ is the present time, $w_o$ is a function which depends only on $w$\footnote{According to the Eqs. \eqref{eq:GLC_fixing}, $w_o$ might also depend on $\tilde\theta^a$. However, as we discuss in the App. \ref{app:w_o}, we choose to avoid this dependency in order to restore the fully non-linear residual gauge freedom and avoid divergent behavior along the observer's geodesic.} and $\chi_o$ and $\hat\chi_o$ depend on $w$ and $\tilde\theta^a$. These free functions reflects exactly the residual gauge freedom already exploited in the GLC gauge \cite{Fanizza:2013doa,Fleury:2016htl,Fanizza:2018tzp}. 
Thanks to this fixing, our set of gauge invariant variables is then given by
\begin{align}
\mathcal{V}=&\nu-\frac{1}{2}D^2\chi-\xi^0 \left( H-\frac{1}{ar} \right)-\frac{\xi^w}{r}\,,\nonumber\\
\mathbb{N}=&N-2H\xi^0+\frac{2}{a}{\xi^0}'-2{\xi^w}'\,,\nonumber\\
\mathcal{M}=&\mu-\chi\,,\nonumber\\
\hat{\mathcal{M}}=&\hat{\mu}-\hat{\chi}\,,\nonumber\\
\mathbb{U}=&u +\frac{1}{ar^{2}}\xi^0-\frac{\xi^w}{r^{2}}-\chi'\,,\nonumber\\
\hat{\mathbb{U}}=&\hat{u}-\hat{\chi}'\,.
\label{eq:GLC_GI_var}
\end{align}

Despite of the fact that these variables are not completely fixed, the great advantage of Eqs. \eqref{eq:GLC_GI_var} is that they corresponds to a linearized version of a fully non-linear gauge choice, which solves exactly the geodesic equation of time-like and light-like particles. Indeed, as above-mentioned, thanks to this properties, physical observables regarding light-like messengers have simple exact expression in such a coordinate system (e.g. angular distance and redshift).

In the following we will explicitly discuss a way to obtain a gauge invariant expression for the angular distance-redshift relation as measured by a free-falling observer, starting from the known result within GLC gauge \cite{Fanizza:2013doa} (see also \cite{BenDayan:2012pp}). This will provide the general gauge invariant expression for this observable and consists of one of the main result presented along this work. Moreover, we will explicitly show how the newly found gauge invariant expression on the  geodesic light-cone coordinate gives a gauge invariant expression in terms of the standard cosmological perturbation theory. Before that, we need to provide the link between the perturbations in Eqs. \eqref{eq:FRWpert} and \eqref{eq:GLCpert}.

\subsection{Map between GLC metric and standard FLRW metric}
The standard perturbation theory of Eqs. \eqref{eq:FRWpert} is related to the perturbative theory on the geodesic light-cone background of Eqs. \eqref{eq:GLCpert} through the finite coordinate transformation of \eqref{eq:finiteCT} as follow
\begin{equation}
\delta{f}_{\mu\nu}=\frac{\partial y^{\alpha}}{\partial x^{\mu}}\frac{\partial y^{\beta}}{\partial x^{\nu}}\delta{g}_{\alpha\beta}\,.
\end{equation}
This leads to the relations
\begin{align}
a^2L=&-2\left(\phi-\frac{1}{2}C_{rr}-\mathcal{B}_r\right)\,,\nonumber\\
aM=&-\left(\mathcal{B}_r+C_{rr}\right)\,,\nonumber\\
N=&C_{rr}\,,\nonumber\\
aV_a=&-\left(\mathcal{B}_a+C_{ra}\right)\,,\nonumber\\
U_a=&C_{ra}\,,\nonumber\\
\delta\gamma_{ab}=&C_{ab}\,,
\label{eq:conditions}
\end{align}
which can be inverted to obtain\footnote{These relations are consistent with the transformations of $\mathcal{B}_r$ and $\mathcal{B}_a$ discussed in \cite{Mitsou:2020czr} for the light-cone coordinates.}
\begin{align}
\phi=&-\frac{1}{2}\left( a^2\,L+N+2aM \right)\,,\nonumber\\
\mathcal{B}_r=& -N-aM\,,\nonumber\\
C_{rr}=& N\,,\nonumber\\
\mathcal{B}_a=&-U_a-aV_a\,,\nonumber\\
C_{ra}=& U_a\,,\nonumber\\
C_{ab}=& \delta\gamma_{ab}\,.
\label{eq:37}
\end{align}

The first line in Eqs. \eqref{eq:37} tells us already an important point. Indeed, from Eqs. \eqref{eq:33} we get that $\xi^0$ can be rewritten as
\beq
\xi^0=\int^\tau_{\tau_{in}}d\tau'\phi(\tau',w-\eta(\tau)+\eta(\tau'))
=\int^\eta_{\eta_{in}}d\eta'\phi(\eta',r)\,.
\label{eq:xi0}
\eeq
This shows that $\xi^0$ is precisely the shift with respect to the time as given in the SG \cite{BenDayan:2012pp}. In particular, this fixing of the time gauge mode implies that all the gauge invariant quantities are established with respect to the time as measured by time-like free-falling rest-frames for a given time $\eta$ and position $r$. In particular, for the angular distance-redshift relation we will show how this term, evaluated along the observer's geodesic, precisely reproduces the observer's time shift already pointed out in \cite{Biern:2016kys} and ensures the gauge invariance of the linearized expression.

Since Eqs. \eqref{eq:conditions} and \eqref{eq:37} holds for any gauge choice, we can combine them with the respective gauge transformations in Eqs. \eqref{eq:FLRW_gauge} and \eqref{eq:GLC_gauge} in order to show the consistency of our relations. In this regard, let us suppose to transform the standard perturbations between two different gauges through Eqs. \eqref{eq:FLRW_gauge}. In the tilded gauge, the same relation in Eqs. \eqref{eq:conditions} holds and allows to find the analogous  geodesic light-cone perturbations in the tilded gauge. As an instance, first of Eqs. \eqref{eq:conditions} must hold in the two gauges
\beq
a^2L=-2\left(\phi-\frac{1}{2}C_{rr}-\mathcal{B}_r\right)\qquad\text{and}\qquad
a^2\widetilde{L}=-2\left(\widetilde{\phi}-\frac{1}{2}\widetilde{C_{rr}}-\widetilde{\mathcal{B}_r}\right)\,.
\label{eq:39}
\eeq
By combining them with Eqs. \eqref{eq:FLRW_gauge} and \eqref{eq:GLC_gauge}, we get that
\beq
a\,\dot{\xi}^w=\partial_\eta\epsilon^\eta
-\partial_{r}\epsilon^r
-\partial_{r}\epsilon^\eta+\partial_\eta\epsilon^r\,,
\label{eq:instance}
\eeq
which is identically satisfied by the transformation rules\footnote{Since, in general, gauge transformations of a tensor $T$ can be seen as the action of the Lie derivative wrt the gauge field, these transformations also follow from the fact that Lie derivative $\mathcal{L}$ itself is independent on the adopted coordinate system, namely $\mathcal{L}_\xi T=\mathcal{L}_\epsilon T$.}
\beq
\xi^\mu=\frac{\pa x^\mu}{\pa y^\nu}\epsilon^\nu\qquad\text{and}\qquad\frac{\pa }{\pa x^\mu}=\frac{\pa y^\nu}{\pa x^\mu}\pa_\nu\,,
\label{eq:tr_rules}
\eeq
applied through the finite coordinate transformation in Eqs. \eqref{eq:finiteCT}. Indeed, Eqs. \eqref{eq:finiteCT} and \eqref{eq:tr_rules} lead to
\bea
\xi^0&=&a\,\epsilon^\eta\qquad,\qquad \xi^w=\epsilon^\eta+\epsilon^r\qquad,\qquad \hat{\xi}^a=\epsilon^a
\nonumber\\
\pa_\tau&=&\frac{1}{a}\left( \pa_\eta-\pa_r \right)\qquad,\qquad\pa_w=\pa_r\qquad\text{and}\qquad\frac{\pa}{\pa\tilde\theta^a}=\frac{\pa}{\pa\theta^a}\,.
\label{eq:intermediate}
\eea
From these relations we get that
\beq
a\,\dot{\xi}^w=\left(\pa_\eta-\pa_r\right)\left( \epsilon^\eta+\epsilon^r \right),
\eeq
which is precisely the r.h.s. of Eq. \eqref{eq:instance} and proves the identity. The other identities can be checked in the same way from the remaining five relations in Eqs. \eqref{eq:conditions}, leading to
\begin{align}
H\xi^0-a\dot{\xi}^w+{\xi^w}'+\dot{\xi}^0=&\partial_{r}\epsilon^\eta-\partial_\eta\epsilon^r
+2\Hcal \epsilon^\eta+2\partial_{r}\epsilon^r\,,\nonumber\\
-2H\xi^0+\frac{2}{a}{\xi^0}'-2{\xi^w}'=&-2\Hcal \epsilon^\eta-2\partial_{r}\epsilon^r\,,\nonumber\\
\partial_{a}\xi^w-a\,\bar{\gamma}_{ab}\dot{\hat{\xi}}^{b}=&
\partial_{a}\epsilon^\eta-\bar{\gamma}^{FRW}_{ab}\partial_\eta\epsilon^{b}
+\partial_{a}\epsilon^r+\bar{\gamma}^{FRW}_{ab}\partial_{r}\epsilon^{b}\,,\nonumber\\
\frac{1}{a}\partial_{a}\xi^0-\partial_{a}\xi^w-\bar{\gamma}_{ab}\,\hat{\xi}^b\,'=&-\partial_{a}\epsilon^r-\bar{\gamma}^{FRW}_{ab}\partial_{r}\epsilon^{b}\,,\nonumber\\
-\frac{1}{a^{2}}\xi^0\left(a^{2}\bar\gamma_{ab}\right)\dot{}
-\xi^w\left(\bar\gamma_{ab}\right)'
-\left(\bar\gamma_{ac}D_{b}+\bar\gamma_{bc}D_{a}\right)\hat{\xi}^{c}=&-\frac{1}{a^{2}}\epsilon^{\rho}\partial_{\rho}\left(a^{2}\bar{\gamma}^{FRW}_{ab}\right)
\nonumber\\
&-(\bar{\gamma}^{FRW}_{ac}\partial_{b}+\bar{\gamma}^{FRW}_{bc}\partial_{a})\epsilon^{c}\,.
\label{eq:314}
\end{align}
Also these relations are identically satisfied when combined with Eqs. \eqref{eq:intermediate}. This proves that the gauge transformations consistently transform between standard perturbations and geodesic light-cone ones.

For later uses, we conclude this section by reporting the following useful relations between the geodesic light-cone perturbations and the standard ones
\bea
D^2v&=&-\frac{1}{ar^2}D^a\left( \mathcal{B}_a+C_{ra} \right)\qquad,\qquad
D^2\hat{v}=-\frac{1}{ar^2}\widetilde{D}^a\left( \mathcal{B}_a+C_{ra} \right)\,,\nonumber\\
D^2u&=&\frac{1}{r^2}D^aC_{ra}\qquad,\qquad
D^2\hat{u}=\frac{1}{r^2}\widetilde{D}^aC_{ra}\,,\nonumber\\
\left(D^2\right)^2\mu&=&\frac{1}{r^2}D^{ab}C_{ab}\qquad,\qquad
\left(D^2\right)^2\hat{\mu}=\frac{1}{r^2}\widetilde{D^{ab}}C_{ab}\qquad\text{and}\qquad
\nu=\frac{1}{4}C^a_a\,.
\label{eq:38}
\eea
Relations in Eqs. \eqref{eq:38} have been obtained by combining the decomposition in Eqs. \eqref{eq:GLC_dec} with Eqs. \eqref{eq:conditions}, thanks to the relations
\bea
q^{ab}D_{(a}D_{b)}&=&D^2\qquad,\qquad
q^{ab}\widetilde{D}_{(a}\widetilde{D}_{b)}=D^2\qquad,\qquad
q^{ab}D_{(a}\widetilde{D}_{b)}=0\,,\\
D^{ab}D_{ab}&=&\frac{1}{2}\left(D^2\right)^2\qquad,\qquad
\widetilde{D}^{ab}\widetilde{D}_{ab}=\frac{1}{2}\left(D^2\right)^2\qquad\text{and}\qquad
\widetilde{D}^{ab}D_{ab}=D^{ab}\widetilde{D}_{ab}=0\,.\nonumber
\eea

Let us conclude this section by commenting Eqs. \eqref{eq:37} and conditions in Eqs. \eqref{eq:GLC_linear_cond}. According to what we have shown between Eqs. \eqref{eq:39} and Eqs. \eqref{eq:314}, we can immediately find the equivalent gauge choice of the linearized GLC gauge in terms of the standard perturbations by imposing Eqs. \eqref{eq:GLC_linear_cond} in Eqs. \eqref{eq:37}. This leads to
\begin{align}
\phi=\,0\qquad,\qquad
\mathcal{B}_r= -\frac{1}{2}C_{rr}\qquad\text{and}\qquad
\mathcal{B}_a=-C_{ra}\,.
\label{eq:OSG}
\end{align}
We note that this choice shares the same condition of the standard SG on the perturbations in $g_{00}$, i.e. $\phi=0$, whereas it requires more involved relations for the $g_{0i}$ metric perturbations (usually in standard SG one imposes $g_{00}=g_{0i}=0$). Since this choice shares the same properties of the GLC gauge, where the angles can be identified with the directions as seen in the observer's rest-frame, we name the new gauge in Eqs. \eqref{eq:OSG} the \textit{Observational Synchronous Gauge} (OSG). However, we note the important difference that the standard SG can be imposed fully non-linearly on the metric components, whereas the non-linear generalization of the OSG requires conditions which will be probably more involved than Eqs. \eqref{eq:OSG}. This is simply due to the fact that Eqs. \eqref{eq:GLC_linear_cond} themselves are a linearization of more involved non-linear gauge conditions. Nevertheless, we will show later that the gauge choice in Eqs. \eqref{eq:OSG} interestingly leads to an expression in standard perturbation theory for the angular distance-redshift relation entirely given in terms of metric perturbations evaluated at the observer and the source position.

\section{Gauge invariant observables in the geodesic light-cone perturbations}
\label{sec:4}
The GLC gauge has shown its advantages in finding exact expressions for light-like observables. Among them, the one we discuss in this work is the angular distance. Indeed this can be derived through the solution of the Jacobi map in terms of the metric \eqref{31} (see \cite{Fanizza:2013doa}). 
As discussed in \cite{Fanizza:2013doa}, the linearized expression for the angular distance itself is a bi-scalar under a gauge transformation and this implies that gauge modes modify the expression of the angular distance for the evaluation both at the observer and source position for any chosen gauge. For what concerns the transformations at the source position, in principle one needs 4 conditions to completely fix this gauge dependence. However, for observables which do not depend on angles at the background level, at linear level it is enough to fix only 2 of them. This is the case of the angular distance. Therefore we need to fix the linear perturbations at the source associated to the time and radial gauge modes, since the background depends only on those coordinates. To this aim, we adopt the observed redshift to fix the time gauge mode and constrain the
radial one by requiring that photons travel on the observed past-light cone, not on the background
one. This renders the angular distance-redshift relation for light-like particles gauge invariant at the source position. 
However, the gauge modes at the observed position must be tackled differently: indeed the time gauge mode at the observer position requires a further fixing, accordingly to what is the motion of the rest frame doing the observations. Only when this choice is done the ultimate expression for the angular distance-redshift relation is completely gauge invariant \cite{Biern:2016kys}.

These properties are completely general, but already show the advantage of adopting the GLC gauge invariant variables, given by Eqs. \eqref{eq:33} and \eqref{eq:GLC_GI_var}, for similar computations. Indeed, the choice of $\xi^w$ in Eqs. \eqref{eq:33} selects the $w$ gauge mode such that the gauge invariant variables coincide with the linearized version of the exact GLC gauge, where $w=\text{constant}$ precisely selects an entire light-like surface at a given time. On the other hand, the choice of $\xi^0$ picks up the constant time hyper-surfaces accordingly to the time measured by a static free-falling time-like flows, as already pointed out after Eq. \eqref{eq:xi0}. Accordingly to what discussed above, this choice for the gauge invariant variables fixes the observer's gauge modes accordingly to what done in \cite{Biern:2016kys}. The only missing thing is then expressing the time gauge mode at the source in terms of the observed redshift. In this regards, GLC properties again simplifies a lot the problem.

Indeed, it has been already shown in \cite{Gasperini:2011us} that the redshift $z$ in the GLC gauge can be easily written as
\beq
1+z=\frac{\Upsilon(\tau_o,w,\tilde\theta^a)}{\Upsilon(\tau,w,\tilde\theta^a)}\,.
\label{eq:GLC_red}
\eeq
This simple expression comes from the fact that the time-like flow, comoving with a geodesic observer, can be written as $u_\mu=\partial_\mu \tau$, whereas the quadri-momentum of a photon, which instead solves the light-like geodesic, can be written as $k_\mu=\partial_\mu w$.

Since we have the non-linear expression for the redshift, we can linearize it by  using the perturbations as expressed in Eq. \eqref{eq:GLCpert} with the conditions \eqref{eq:GLC_linear_cond}. This leads to
\beq
1+z=\frac{a_o}{a_s}\left[1-\left(aM\right)|^o_s\,\right]=\frac{a_o}{a_s}\left(1+\frac{1}{2}\,N|^o_s\right)\,,
\label{eq:42}
\eeq
where the letters $o$ and $s$ respectively indicate quantities evaluated at the observer and at the source time and the notation $N|^o_s\equiv N_o-N_s$.

Whereas in cosmology the proper time in our rest frame can be set as the reference time at the observer position, the one in the source rest frame cannot be measured. On the other hand, since redshift from a given source is easily accessible, it turns out to be very suitable to adopt the redshift $z$ rather than $\tau$ for the comparison with data.
Hereafter, we then call $\tau_z$ the proper time evaluated at the observed redshift. The two spheres at $\tau=\text{constant}$ and $\tau_z=\text{constant}$ are equal on the background,  but differ already at linear order because of the perturbations in Eq. \eqref{eq:42}. We can then parameterize this linear difference with
\beq
\delta\tau_z\equiv\tau-\tau_z\,,
\label{eq:tau_z}
\eeq
and using Eq. \eqref{eq:42} we can evaluate this time shift as follow. Following \cite{Fanizza:2015swa}\footnote{See also \cite{Challinor:2011bk,Jeong:2014ufa,Jeong:2013psa,Jeong:2011as} for similar evaluation in the framework of standard perturbation theory rather than GLC one.}, we first linear expand $a(\tau)=a(\tau_z)\left[1+H(\tau_z)\delta\tau_z\right]$ in Eq. \eqref{eq:42} and get
\beq
1+z=\frac{a_o}{a(\tau_z)}\left(1-H_z\delta\tau_z+\frac{1}{2}\,N|^o_s\right)\,,
\label{eq:tauz}
\eeq
where we have defined $H_z\equiv H(\tau_z)$. Then we require the parenthesis in Eq. \eqref{eq:tauz} to be equal to $1$. As a consequence, we have
\beq
\delta\tau_z=\frac{1}{2 H_z}\,N|^o_s\,.
\label{eq:redshift_shift}
\eeq
The result in Eq. \eqref{eq:redshift_shift} allows to write a generic scalar observable $\mathcal{O}$ in terms of the observed redshift. Let us show this procedure in the simple, illustrative case when the background observable $\bar{\mathcal{O}}$ depends only on $\tau$.  At linear level, we can write
\beq
\mathcal{O}=\bar{\mathcal{O}}(\tau)+\delta\mathcal{O}(\tau,w,\tilde\theta^a)\,,
\label{eq:observable}
\eeq
where $\delta\mathcal{O}$ are the linear perturbations. From Eq. \eqref{eq:tau_z}, we can then expand $\bar{\mathcal{O}}$ in Eq. \eqref{eq:observable} and obtain
\beq
\bar{\mathcal{O}}(\tau)+\delta\mathcal{O}(\tau,w,\tilde\theta^a)
=\bar{\mathcal{O}}(\tau_z)+\delta\mathcal{O}(\tau_z,w,\tilde\theta^a)+\delta\tau_z\dot{\bar{\mathcal{O}}}
=\bar{\mathcal{O}}(\tau_z)+\delta\mathcal{O}_z(\tau_z,w,\tilde\theta^a)
\label{eq:49}
\eeq
where we have defined $\delta\mathcal{O}_z(\tau_z)=\delta\mathcal{O}+\delta\tau_z\dot{\bar{\mathcal{O}}}$ which indeed represents the linear perturbation for the observable $\mathcal{O}$ in terms of the observed redshift.

In particular, since a plethora of observables can be exactly written in the GLC gauge, any of them can be expressed in terms of the observed linear redshift thanks to the procedure presented so far. We can then infer the linear gauge invariant expression for such observables in the new set of geodesic light-cone perturbations. To this aim we use Eqs. \eqref{eq:GLC_GI_var} and adopt the following procedure
\begin{itemize}
\item First we identify the chosen observable whose expression can be exactly written in the non-linear GLC gauge.
\item Then we linearize this quantity in order to have it in terms of the linear perturbations $\left( N,u,\hat{u},\nu,\mu,\hat\mu \right)$.
\item Hence, we express this linearized expression in term of the observed redshift accordingly to the procedure presented above.
\item Finally we promote the linear perturbations as their appear in the observable to their gauge invariant counterparts from Eqs. \eqref{eq:GLC_GI_var}, i.e. $\left( N,u,\hat{u},\nu,\mu,\hat\mu \right)\rightarrow\left( \mathbb{N},\mathbb{U},\hat{\mathbb{U}},\mathcal{V},\mathcal{M},\hat{\mathcal{M}} \right)$.
\end{itemize}
This procedure automatically provide gauge invariant expressions for any observable in terms of the observed redshift by construction. Moreover, it preserves the advantage of the GLC gauge where light-like observables have simple expressions. In the following section, we will apply this procedure to the angular distance-redshift relation and show that the obtained result is indeed gauge invariant also once expressed in terms of standard cosmological perturbations.

Before concluding this section, we would like to underline the following remark. After the substitution with gauge invariant variables $N\rightarrow\mathbb{N}$, the effect in Eq. \eqref{eq:redshift_shift} becomes
\begin{equation}
\delta\tau_z\rightarrow\delta\mathcal{T}_z\equiv \frac{1}{2 H_z}\,\mathbb{N}|^o_s\,.
\end{equation} 
This term has been independently derived also with other approaches (see \cite{Challinor:2011bk,Jeong:2014ufa,Jeong:2013psa,Jeong:2011as}). In this regard, we stress that $\delta\mathcal{T}_z$ coincides with the linearized version of Eq. (8) in \cite{Jeong:2013psa}. Moreover, it is gauge invariant by construction and this is in agreement with its interpretation given in Sect. IIb of \cite{Jeong:2013psa}. The advantage of our procedure consists of taking into account by construction the perturbation of the scale factor discussed in Sect. IIb of \cite{Jeong:2013psa}. In our formalism, this is nothing but $H_o\,\xi^0_o$ contained in $\mathbb{N}_o$. This can be easily seen by noticing that the first equation of Eqs \eqref{eq:conditions} combined within the first equation of Eqs. \eqref{eq:33} precisely returns Eq. (13) of \cite{Jeong:2013psa}.

\section{Angular distance-redshift relation}
\label{sec:5}
As above-mentioned GLC gauge has shown its great advantage also in the evaluation of the angular distance $d_A$. Indeed, its fully non linear expression can be written in this gauge as \cite{Fanizza:2013doa}
\beq
d^2_A=\frac{\sqrt{\gamma}}{\left(\frac{\text{det}\dot\gamma_{ab}}{4\,\sqrt{\gamma}}\right)_o}\,,
\label{eq:GLC_dA}
\eeq
where $\gamma=\text{det}\gamma_{ab}$. Let us now follow the prescription outlined at the end of Sect. \ref{sec:4}. In order to linearize Eq. \eqref{eq:GLC_dA}, we firstly note that
\beq
\gamma=\text{det}\left[a^2 r^2 q_{ab}\left( 1+2 \nu \right)\right]
=a^4 r^4 \sin^2\tilde\theta^1\left( 1+4 \nu \right)\,,
\eeq
since $D_{ab}\mu+\widetilde{D}_{ab}\hat\mu$ are traceless. On the other hand,
\beq
\dot\gamma_{ab}=2\,ar\left\{q_{ab}\left[\left(aHr-1\right)\left( 1+ 2\nu\right)+ar\dot\nu\right]
+\left(aHr-1\right)\left(D_{ab}\mu+\widetilde{D}_{ab}\hat\mu\right)
+\left(D_{ab}\dot\mu+\widetilde{D}_{ab}\dot{\hat\mu}\right)\right\}\,,
\eeq
so that
\beq
\text{det}\dot\gamma_{ab}=4 a^2 r^2\sin^2\tilde\theta^1\left( aHr-1 \right)^2\left[ 1+4\nu+2 \frac{a r}{aHr-1}\dot{\nu} \right]\,.
\eeq
We then easily get that the denominator in Eq. \eqref{eq:GLC_dA} is
\beq
\left(\frac{\text{det}\dot\gamma_{ab}}{4\,\sqrt{\gamma}}\right)_o
=\sin\tilde\theta^1\,\lim_{\tau\rightarrow\tau_o}\left[\left( aHr-1 \right)^2\left( 1+2\nu+2 \frac{a r}{aHr-1}\dot{\nu} \right)
 \right]\,.
 \label{eq:GLC_dA_den_lim}
\eeq
For the moment, we leave unresolved the limit in Eq. \eqref{eq:GLC_dA_den_lim} and write the linearized version of the angular distance \eqref{eq:GLC_dA} as
\beq
d_A=ar\frac{\left( 1+\nu \right)}{\lim_{\tau\rightarrow\tau_o}\left[\left( aHr-1 \right)\left( 1+\nu+ \frac{a r}{aHr-1}\dot{\nu} \right)
 \right]}\,.
 \label{eq:56}
\eeq
Using the result obtained for $\delta\tau_z$ in Eq. \eqref{eq:redshift_shift}, we can then find the desired expression for the angular distance-redshift relation. In particular, from Eq. \eqref{eq:56}, we expand the background terms as
\beq
a(\tau)r(\tau)=a_zr_z\left[ 1+\left(H_z-\frac{1}{a_zr_z}\right)\delta\tau_z \right]
=a_zr_z\left[ 1+\frac{1}{2}\left(1-\frac{1}{a_zH_zr_z}\right)\,N|^o_z \right]\,,
\label{eq:dA_redshift_corr}
\eeq
where the subscript $z$ reminds that the quantities are evaluated at $\tau_z$. At this point, the sum of Eqs. \eqref{eq:56} and \eqref{eq:dA_redshift_corr} gives
\beq
d_A=ar\frac{\left[ 1+\nu -\frac{1}{2}\left(1-\frac{1}{a_zH_zr_z}\right)\,N|^z_o\right]}{\lim_{\tau\rightarrow\tau_o}\left[\left( aHr-1 \right)\left( 1+\nu+\frac{a r}{aHr-1}\dot{\nu} \right)
 \right]}\,.
\eeq

Hence, according to our prescription, the gauge invariant expression for the angular distance-redshift relation in terms of the geodesic light-cone perturbations is
\beq
d_A=ar\frac{\left[ 1+\mathcal{V}-\frac{1}{2}\left(1-\frac{1}{a_zH_zr_z}\right)\,\mathbb{N}|^z_o \right]}{\lim_{\tau\rightarrow\tau_o}\left[\left( aHr-1 \right)\left( 1+\mathcal{V}+\frac{a r}{aHr-1}\dot{\mathcal{V}} \right)
 \right]}\,.
 \label{eq:GLC_unresolved_GI_1}
\eeq
The reason why we left unresolved the limit stands in the fact that the substitution $\nu\rightarrow\mathcal{V}$ introduces in the expression terms scaling as $r^{-1}$ (see Eqs. \eqref{eq:GLC_GI_var}). These terms disappear in the GLC result but contribute to the total gauge invariant result. In particular, we have that
\beq
\dot{\mathcal{V}}=\dot{\nu}
-\frac{1}{2}D^2v
-H\dot{\xi}^0-\xi^0\dot{H}
+\frac{1}{r}\dot{\left(\frac{\xi^0}{a}-\xi^w\right)}
+\frac{1}{ar^2}\left(\frac{\xi^0}{a}-\xi^w-\frac{1}{2}D^2\xi^w\right)\,.
\label{eq:414}
\eeq
The situation looks even worst than expected since we have also terms diverging as $r^{-2}$. However, this terms can be canceled thanks to the free function $w_o$ in $\xi^w$. Indeed the requirement that $\xi^w_o=\frac{\xi^0_o}{a_o}$ cancels this divergence and fixes $w_o=\frac{\xi^0_o}{a_o}$. On the contrary,  let us note that the time derivative of $\xi^0/a-\xi^w$ at the observer time is not null. This is due to the fact that $w_o$ is independent on $\tau$. In this way, it can not erase $\dot\xi^0$ at $\tau_o$. This is a subtle but very important point since, as we will show later on, the time derivative of $\xi^0$ is crucial to obtain all the terms which reconstruct the observer velocity in the final expression of the angular distance. We also note that, from Eqs. \eqref{eq:38}, $D^2 v$ contains in principle $r^{-2}$ terms, when expressed in terms of the standard perturbations. However, we assume that these terms cancels out along the observer's geodesic since they are sourced either by vector perturbations or angular derivatives of scalar fields. In fact, any angular dependence should not appear along the observer's geodesic to avoid multi-valued functions on the observer's world-line.

Given that we can rewrite the limit in the denominator of Eq. \eqref{eq:GLC_unresolved_GI_1} in a more compact form as
\beq
d_A=ar\frac{\left[ 1+\mathcal{V} -\frac{1}{2}\left(1-\frac{1}{a_zH_zr_z}\right)\,\mathbb{N}|^z_o\right]}{\left( 1+\mathcal{V}- a r\dot{\mathcal{V}} \right)_o}\,,
 \label{eq:GLC_dA_redshift_GI}
\eeq
where we have used the subscript $o$ for the limit $\tau\rightarrow\tau_o$. The result in Eq. \eqref{eq:GLC_dA_redshift_GI} represents the gauge invariant expression of the angular distance-redshift relation in terms of the proper time of the free-falling observer static in the SG. This is one of the most important result in this work.

In the following section, we will discuss the agreement between Eq. \eqref{eq:GLC_dA_redshift_GI} and the gauge invariant result as written in terms of the standard cosmological perturbations already obtained in \cite{Scaccabarozzi:2017ncm}. Before that, we conclude this section by commenting on the appearance of $\left(ar\dot{\mathcal{V}}\right)_o$. From Eq. \eqref{eq:414} we obtain
\bea
\left( 1+\mathcal{V}-a r\dot{\mathcal{V}} \right)_o
&=& \left(1+\nu-\frac{1}{2}D^2\chi
-\dot{\xi}^0
+a\dot{\xi}^w
-ar\,\dot{\nu}
\right)_o\nonumber\\
&=&\left(1+\nu
+\frac{N}{2}+aM
-\frac{1}{2}D^2\chi
+\frac{{\xi^0}'}{a}
-ar\,\dot{\nu}
\right)_o\,,
\label{eq:421}
\eea
where in the last line we have used Eqs. \eqref{eq:GLC_fixing}.
We underline that the term ${\xi^0_o}'$ represents the observer peculiar velocity. Indeed, a generic function $Q$ changes according to the background transformation between standard and geodesic light-cone coordinates as
\bea
Q(\tau',w-\eta(\tau)-\eta(\tau)')'&=&\pa_rQ(\eta',\eta_o-\eta)\\
Q(\tau',w)'&=&\pa_rQ(\eta',\eta_o-\eta')
=\pa_{\eta'}Q(\eta',\eta_o-\eta')-\frac{d}{d\eta'}Q(\eta',\eta_o-\eta')\nonumber\,.
\label{eq:513}
\eea
In this way, we get that
\beq
\frac{{\xi^0_o}'}{a_o}=-\int^{\eta_o}_{\eta_{in}}d\eta'\frac{a(\eta')}{a(\eta_o)}\pa_r\phi(\eta')\,,
\eeq
which precisely reproduces the peculiar velocity of a free falling observer. The presence of this term can be interpreted as the aberration of a solid angle as seen by a moving observer predicted by special relativity. This is in agreement with the fact that the denominator is nothing but the solid angle subtended by source as seen by the observer. We redirect the interested reader to \cite{Fanizza:2013doa} for the complete discussion and evaluation of this effect in the non-linear LG (also mentioned as Poisson Gauge). We remark that our linear result in Eq. \eqref{eq:421} is in agreement with the one presented in \cite{Fanizza:2013doa}.

\section{Gauge invariance and comparison with literature}
\label{sec:6}
To conclude, we now discuss the agreement between the result in Eq. \eqref{eq:GLC_dA_redshift_GI} with the standard linear expression for the gauge invariant angular distance-redshift relation (see \cite{Scaccabarozzi:2017ncm}). To this aim, we first combine Eq. \eqref{eq:GLC_dA_redshift_GI} with Eqs. \eqref{eq:GLC_GI_var} and \eqref{eq:421} to obtain
\bea
d_A(z)&=&a_zr_z\left[ 1
+\nu_z-\frac{1}{2}D^2\chi_z-\frac{\xi_z^w}{r_z}
-\left(1-\frac{1}{a_zH_zr_z}\right)\frac{{\xi_z^0}'}{a_z}
-\frac{1}{2}\left(1-\frac{1}{a_zH_zr_z}\right)\,
\left( N-2{\xi^w}' \right)^z_o \right.\nonumber\\
&&\left.-\frac{1}{a_zH_zr_z}\frac{{\xi^0_o}'}{a_o}
-\left(1-\frac{1}{a_zH_zr_z}\right)H_o\xi_o^0
-\nu_o
-\frac{N_o}{2}-a_oM_o
+\frac{1}{2}D^2\chi_o
+\left(ar\dot{\nu}\right)_o
\right]\,.
\label{eq:51}
\eea
As anticipated, let us note that the time shift $\xi^0$ at the source point disappeared. This reflects the fact that we have used the redshift to parametrize the observed age of the source. Moreover, the aforementioned velocity aberration of the observed solid angle breaks the symmetry between the source and the observer peculiar velocity. This feature is also present in the standard linear expression for $d_A(z)$. Let us now explicitly show in
Eq. \eqref{eq:51} the perturbations on our geodesic light-cone background, using Eqs. \eqref{eq:33}  with the regularity condition $w_o=\xi^0_o/a_o$. We then obtain
\bea
d_A(z)&=&a_zr_z\left[ 1
+\nu_z+\frac{1}{2}\int^{\tau_o}_{\tau_z} d\tau' \left(D^2v +\frac{1}{2\,ar^{2}}\int^{\tau_o}_{\tau'} d\tau''\,a D^2L\right)\right.\nonumber\\
&&-\frac{1}{2\,r_z}\int^{\tau_o}_{\tau_z} d\tau'\,a L
+\frac{1}{2}\left(1-\frac{1}{a_zH_zr_z}\right)\frac{1}{a_z}\int^{\tau_z}_{\tau_{in}}d\tau'\left(N+2aM+a^2\,L\right)'\nonumber\\
&&-\frac{1}{2}\left(1-\frac{1}{a_zH_zr_z}\right)\,N^z_o
+\frac{1}{2}\left(1-\frac{1}{a_zH_zr_z}\right)\,\int^{\tau_o}_{\tau_z} d\tau'\,a L'\nonumber\\
&&+\frac{1}{2\,a_zH_zr_z}\frac{1}{a_o}\int^{\tau_o}_{\tau_{in}}d\tau'\left(N+2aM+a^2\,L\right)'
-\nu_o
-\frac{1}{2}N_o-a_oM_o
+\left(ar\dot\nu \right)_o\nonumber\\
&&\left.+\frac{1}{2}\left(H_o-\frac{H_o}{a_zH_zr_z}+\frac{1}{a_or_z}\right)\int^{\tau_o}_{\tau_{in}}d\tau'\left(N+2aM+a^2\,L\right)
\right]\,,
\label{eq:GLC_GI_dA_z}
\eea
where we remind that the time integral within $\xi^0$ is done along the curve $w-\eta(\tau)+\eta(\tau')$. Eq. \eqref{eq:GLC_GI_dA_z} represents the gauge invariant expression for the angular distance-redshift relation in terms of the geodesic light-cone perturbations. From Eq. \eqref{eq:GLC_GI_dA_z} and the conditions in Eqs. \eqref{eq:GLC_linear_cond}, we immediately note that the absence of integrated terms in the linear angular distance-redshift relation is a peculiar feature of the GLC gauge. This simply reflects the factorization between source and observer terms in the exact solution found in \cite{Fanizza:2013doa}. However, other gauge choices admit integrated terms even in terms of light-like perturbations.

In particular, we can have a better understanding about the integrated terms in Eq. \eqref{eq:GLC_GI_dA_z} through Eqs. \eqref{eq:conditions}, \eqref{eq:37} and \eqref{eq:38}. If we transform the integrated terms according to the finite coordinate transformations of Eqs. \eqref{eq:finiteCT} and use the result  in Eq. \eqref{eq:513}, we finally obtain
\bea
&&\int^{\tau_o}_\tau d\tau'\,a L'=-2\int^{\eta_o}_\eta d\eta'\,\pa_{\eta'}\left(\phi-\frac{1}{2}C_{rr}-\mathcal{B}_r\right)
+2\left.\left(\phi-\frac{1}{2}C_{rr}-\mathcal{B}_r\right)\right|^o_s\,,\nonumber\\
&&\int^{\tau_o}_\tau d\tau'\,a L=-2\int^{\eta_o}_\eta d\eta'\,\left(\phi-\frac{1}{2}C_{rr}-\mathcal{B}_r\right)\,,\nonumber\\
&&\int^{\tau_o}_{\tau_z} d\tau' \left(D^2v +\frac{1}{2\,ar^{2}}\int^{\tau_o}_{\tau'} d\tau''\,a D^2L\right)=
-\int^{\eta_o}_{\eta_z} \frac{d\eta'}{r^2} \left[D^a\left( \mathcal{B}_a+C_{ra} \right) \right.\nonumber\\
&&\left.+\int^{\eta_o}_{\eta'} d\eta''\, D^2\left(\phi-\frac{1}{2}C_{rr}-\mathcal{B}_r\right)\right]\,,\nonumber\\
&&\int^{\tau_z}_{\tau_{in}}d\tau'\left(N+2aM+a^2\,L\right)'=-\frac{1}{a_z}\int^{\eta_z}_{\eta_{in}}d\eta'a\pa_r\phi\,.
\label{eq:integrals}
\eea
We then recognize that  first integral in Eq. \eqref{eq:integrals} reproduces local and integrated Sachs-Wolfe effect, whereas the second line corresponds to the time-delay. Finally third and last equality respectively reproduces lensing effects and peculiar velocity. We also remark that all the r.h.s in Eqs. \eqref{eq:integrals} are null in the OSG provided in Eqs. \eqref{eq:OSG}.

Before concluding our comparison, we briefly comment on three aspects. First of all, Eq. \eqref{eq:GLC_GI_dA_z} does not contain neither $\chi_o$ nor $\hat\chi_o$. These residual gauge modes are fixed in order to require the observational gauge, where the GLC angles are chosen to coincide with the observed direction of the incoming photons in the observer rest frame \cite{Fanizza:2018tzp}. This independence has its origin in the fact that the background angular distance does not depend on the angle. Secondly, we note that the term $D^2w_o$, coming from $D^2\chi$, in Eq. \eqref{eq:51} is not present. This is due to the fact that the residual degree of freedom is set to be equal to $\xi^0_o$. The latter contribute only as a monopole and then its eigenvalue on the Laplacian is $0$. This implies then $D^2w_o=0$. Another way to see this is simply that the fully non-linear residual gauge freedom of the GLC gauge does not admit any angular dependence on $w_o$, so that any angular derivative acting on it is null. Finally, we stress that the last term in Eq. \eqref{eq:GLC_GI_dA_z} is exactly the time lapse already exploited in the literature. This can be trivially seen by recalling Eq. \eqref{eq:37}.

At this point, we can express Eq. \eqref{eq:GLC_GI_dA_z} in terms of standard cosmological perturbations through Eqs. \eqref{eq:conditions}, \eqref{eq:37} and \eqref{eq:38}. Moreover, we also change the integrated terms according to Eqs. \eqref{eq:integrals}. In this way, Eq. \eqref{eq:GLC_GI_dA_z} gives
\bea
d_A(z)&=&a_zr_z\left\{ 1
+\frac{1}{4}{C^a_a}_z-\frac{1}{2}\int^{\eta_o}_{\eta_z} \frac{d\eta'}{r^2} \left[D^a\left( \mathcal{B}_a+C_{ra} \right) +\int^{\eta_o}_{\eta'} d\eta''\, D^2\left(\phi-\frac{1}{2}C_{rr}-\mathcal{B}_r\right)\right]\right.\nonumber\\
&&+\frac{1}{r_z}\int^{\eta_o}_{\eta_z} d\eta'\,\left(\phi-\frac{1}{2}C_{rr}-\mathcal{B}_r\right)
-\left(1-\frac{1}{\mathcal{H}_zr_z}\right)\frac{1}{a_z}\int^{\eta_z}_{\eta_{in}}d\eta'a\pa_r\phi\nonumber\\
&&-\left(1-\frac{1}{\mathcal{H}_zr_z}\right)\,\left[ \int^{\eta_o}_{\eta_z} d\eta'\,\pa_{\eta'}\left(\phi-\frac{1}{2}C_{rr}-\mathcal{B}_r\right)
+\left(\phi-\mathcal{B}_r\right)^z_o \right]\nonumber\\
&&-\frac{1}{a_o\mathcal{H}_zr_z}\int^{\eta_o}_{\eta_{in}}d\eta'a\pa_r\phi
+\left[\mathcal{B}_{r\,o}+\frac{1}{2}C_{rr\,o}-\frac{1}{4}{C^a_a}_o+\frac{1}{4}\left(r\pa_\eta C^a_a-r\pa_r C^a_a\right)_o\right]\nonumber\\
&&\left.-\frac{1}{a_o}\left(\mathcal{H}_o-\frac{\mathcal{H}_o}{\mathcal{H}_zr_z}+\frac{1}{r_z}\right)\int^{\eta_o}_{\eta_{in}}d\eta'a\phi
\right\}\,.
\label{eq:66}
\eea
The result in Eq. \eqref{eq:66} can be combined with Eqs. \eqref{eq:FLRW_gauge} in order to check that $d_A(z)$ is indeed gauge invariant as expected. After a long but straightforward evaluation, we get that
\bea
\widetilde{d_A}(z)&=&d_A(z)\left[ 1
-\left.\frac{1}{2}\left( \frac{1}{r} \right)\right|^o_zD^2\left( \epsilon^\eta+\epsilon^r \right)_o
-\frac{1}{r_z}\left(\epsilon^r\right)_o\right]\,,
\label{eq:GR}
\eea
which explicitly shows that the expression is gauge invariant at the source position. Moreover, there are no pure time gauge modes at the observer position. This is precisely due to the presence of the time lapse at the observer position correctly taken into account. For what concerns the Laplacian and the radial gauge mode left, we can show that they are null as a consequence of the choice $w_0=\xi^0_o/a_o$. Indeed, thanks to the finite coordinate transformation \eqref{eq:finiteCT}, we have that (details are provided in Appendix \ref{app:w_o})
\beq
\epsilon^\eta=\frac{\xi^0}{a}\qquad\text{and}\qquad
\epsilon^r=-\frac{\xi^0}{a}+\xi^w\,.
\eeq
Recalling Eqs. \eqref{eq:33}, in the limit $\tau\rightarrow\tau_o$ we then obtain that
\beq
-\left.\frac{1}{2}\left( \frac{1}{r} \right)\right|^o_zD^2\left( \epsilon^\eta+\epsilon^r \right)_o
-\frac{1}{r_z}\left(\epsilon^r\right)_o
=
-\left.\frac{1}{2}\left( \frac{1}{r} \right)\right|^o_zD^2 w_o
+\frac{1}{r_z}\left(\frac{\xi^0_o}{a_o}-w_o\right)\,,
\eeq
which is indeed null since $w_o$ is a monopole (see Appendix \ref{app:w_o}). Hence this confirms that Eq. \eqref{eq:GLC_GI_dA_z} is the gauge invariant expression for the angular distance-redshift relation in terms of the geodesic light-cone perturbations. We also recall that this condition is necessary to cancel possible divergent behavior of the perturbations around the observer position.

To conclude, let us comment on our results in Eqs. \eqref{eq:66} and \eqref{eq:GR}. The gauge invariance of the angular distance-redshift relation is ensured once the complete fixing of the residual gauge freedom in $\xi^w$ is provided. In particular, we get that the same condition erases at the same time second and third terms in Eq. \eqref{eq:GR}. This result is in contrast with what obtained in \cite{Scaccabarozzi:2017ncm}, where the term $\epsilon^r_o$ is canceled by a specular term coming from the divergence of the linear deflection angles (called $\kappa$ in \cite{Scaccabarozzi:2017ncm}) and it is then claimed that no fixing for $w_o$ is needed for the gauge invariance. In our formalism, the analogous of $\kappa$ in \cite{Scaccabarozzi:2017ncm} is precisely the last term in the first line of Eq. \eqref{eq:66}. However, we are not able to achieve the cancellation. The reason for that stands in the fact that the linear deflection angles involves only angular modes for the gauge transformation (see Appendix A of \cite{Fanizza:2018qux}). This then implies that its angular divergence simply transforms as the angular divergence of the angular gauge modes ($D^2\chi$ in our formalism) which indeed cancels in our derivation with the analogous term in Eqs. \eqref{eq:421} and \eqref{eq:51}, respectively at the observer and source position. From a technical point of view, we believe that the difference wrt \cite{Scaccabarozzi:2017ncm} is due to the following aspect: in Appendix B of \cite{Scaccabarozzi:2017ncm} $w_o$ is fixed such that it exhibits dependence also on angular coordinates. This freedom is not admitted according to the exact residual gauge freedom allowed by the GLC gauge. This is the reason why in this work we have preferred to take $w_o$ independent on the angular coordinates. On the other hand, the gauge modes concerning $\chi_o$ and $\hat\chi_o$ cancels in the final expression. This is in agreement with \cite{Scaccabarozzi:2017ncm}.

\section{Discussion and conclusion}
\label{sec:7}
In this paper we have presented a new linear perturbation theory built on background geodesic light-cone coordinates. Therefore, we have considered a set of coordinates which shares the same properties of the Geodesic Light-Cone (GLC) gauge \cite{Gasperini:2011us} only at the background level. More specifically, we have chosen the time-like coordinate to coincide with the cosmic time and the spatial coordinate to be null, such that it can label different unperturbed past light-cones. On top of this background, we have added linear perturbations without any gauge fixing, with the aim of studying the gauge transformation properties of such perturbations.

After that, we have found how this new set of perturbations is linked to standard perturbation theory, and provided their gauge transformations as generated by infinitesimal linear diffeomorphisms. In this way, after a well-suited decomposition of both perturbations and gauge modes in terms of the rotations on the unitary sphere (i.e., in term of a background symmetry), we have exploited several interesting gauge fixing. Moreover, we have used the values of the perturbations in these gauges to build gauge invariant variables case by case.

After this general treatment, we have provided the derivation of gauge invariant variables such that their values are the same as the ones in the linearized GLC gauge. This choice indeed turns out to be very useful to study in full generality the gauge transformations of physical observables and, at the same time, preserve the simplicity of the GLC gauge for what concerns the study of light-like observables. Furthermore, in Sect. \ref{sec:3} we have also obtained the gauge in standard perturbation theory which corresponds to the linearized GLC gauge and, as a consequence, shares its property:  the proper time coincides with the one measured by a free-falling observer and the angles can be identified with the directions as seen in the observer's rest-frame. We have dubbed this new gauge  \textit{Observational Synchronous Gauge}, since this is a more connected to observations alternative to the standard Synchronous Gauge (SG).

In regard of the derivation of gauge invariant variables, as an example, we have applied the procedure to the case of the angular distance-redshift relation and provided its general expression in terms of the new light-cone perturbations. We stress that our procedure provides an expression for any light-like observables which is gauge invariant both at the observer and the source position. In particular, the term which guarantees the gauge invariance for temporal gauge transformation at the observer position precisely coincides with the one exploited in literature \cite{Biern:2016kys}. Its physical origin is due to the time shift between a free-falling static observer in the SG and a moving geodetic one in any other gauge. Since the coordinate time in the GLC gauge is fully non-linearly build to coincide with the time of a free-falling observer static in the SG, it is not surprising that our result furnish the expected result.

Finally, we have written the obtained expression in terms of standard perturbation theory. In this way, we have verified that our result is gauge invariant also in terms of the standard perturbations, once a partial fixing of the light-like gauge modes is provided. The condition we need to impose can be seen in standard perturbation theory as the geometrical requirement that the center of coordinates coincides with the observer position. Indeed, as shown in Appendix \ref{app:w_o}, a different choice would lead to a dipolar correction at the observer position and this is precisely what one would expect in case of such a misalignment.

The fact that we need to partially fix the residual gauge freedom of the GLC gauge is not surprising. The same happens also in standard perturbation theory when we attempt to evaluate the angular distance within the SG. In that case, it has already been shown that the complete fixing of the residual spatial gauge modes typical of the SG is necessary in order to reproduce the expected gauge transformation of the angular distance to the longitudinal gauge  \cite{Fanizza:2013doa}.

To conclude, we stress that the results obtained in this paper are a crucial step towards a cosmological perturbation theory entirely solved within a set of geodesic light-cone coordinates. 
In fact, on one side, 
since physical light-like observables have gauge invariant expressions their cosmological dynamics can be evolved for any gauge fixing.
On the other hand, the next step will be to develop our formulation for gauge invariant observables to the next-to-leading order in perturbation theory. Especially for what concerns the angular distance-redshift relation, indeed, the full control about observer terms at second order in perturbation theory will finally provide the full gauge invariant expression for the $d_A(z)$. In fact, in this regard, it has been already shown that source and integrated terms at second order can be derived in two different ways, leading to the same results (in \cite{Fanizza:2015swa}, this has been shown within the Poisson Gauge). The addition of the full set of second-order observer's terms will finally provide the entire non-linear second order angular distance-redshift relation. We plan to achieve this result in a forthcoming work.

\section*{Acknowledgement}
GM and MM are supported in part by INFN under the program TAsP ({\it Theoretical Astroparticle Physics}). GF acknowledges support by FCT under the program {\it Stimulus} with the grant no. CEECIND/04399/2017/CP1387/CT0026.


\appendix

\section{Fixing $w_o$}
\label{app:w_o}
Here, we discuss how to derive solution for $w_o$ in Eqs. \eqref{eq:33}. As anticipated, to do so we require that the contribution from $w_o$ is such that to cancel the divergencies associated to the negative powers of $r$ in Eqs. \eqref{eq:GLC_GI_var} for $\mathcal{V}$, i.e.
\beq
\left(1+\frac{D^2}{2}\right)w_o=\frac{\xi^0_o}{a_o}\,.
\eeq
Since $\xi^0_o$ is a monopole, the latter equation is solved in general by the solution
\beq
w_o=\frac{\xi^0_o}{a_o}+d_o(w)P_1\left( \cos\tilde\theta^1 \right)\,,
\label{eq:B2}
\eeq
where $d_o$ is a free function of $w$ and $P_1$ is the Legendre polynomial of order 1, namely $d_o(w)$ is a pure dipole. The presence of an unspecified dipole in the residual gauge mode of $w$ is a peculiar feature of the linearized GLC gauge. Indeed, the non-linear GLC gauge only admits a residual gauge freedom on $w$ which does not depend on $\tilde\theta^a$, namely $w\rightarrow W(w)$ \cite{Fanizza:2013doa,Fleury:2016htl,Mitsou:2017ynv,Fanizza:2018tzp}. Since the non-linear symmetry is stronger than the linear one, we require that the non-linear symmetry must be satisfied also at the linear level and this leads to $d_o(w)=0$. Hence we get
\beq
w_o=\frac{\xi^0_o}{a_o}\,.
\label{eq:B3}
\eeq
This condition has the following meaning. We have that, thanks to Eq. \eqref{eq:finiteCT}, one can write
\beq
\epsilon^r=\frac{\pa\,r}{\pa x^\mu} \xi^\mu
=\frac{\pa\,r}{\pa \tau} \xi^0
+\frac{\pa\,r}{\pa w} \xi^w
=- \frac{\xi^0}{a}+\xi^w\,,
\eeq
and this then implies that at the observer position we get
\beq
\epsilon^r_o=-\frac{\xi^0_o}{a_o}+w_o\,,
\eeq
which is indeed null thanks to Eq. \eqref{eq:B3}. Hence, we conclude that the residual gauge fixing in Eq. \eqref{eq:B3} is equivalent to require that the radial gauge modes at the observer is null. In other words, we impose that the origin of the polar coordinates always coincide with the observer's position. This conclusion is supported from the fact that a different choice from Eq. \eqref{eq:B3} would introduce a dipole in the gauge fixing (just as pointed out in Eq. \eqref{eq:B2}), which is precisely the same kind of angular dependence one would get in case the observer and the coordinate system were misaligned.

\bibliographystyle{JHEP}
\bibliography{biblio}

\end{document}